\documentclass[5p]{elsarticle}
\usepackage[utf8]{inputenc}
\usepackage[english]{babel}
\usepackage{amsmath}
\usepackage{amsfonts}
\usepackage{amssymb}
\usepackage{makeidx}
\usepackage{graphicx}
\usepackage{color}
\biboptions{sort&compress}
\newcommand{\ket}[1]{|#1\rangle}
\newcommand{\bra}[1]{\langle #1|}

\def\cH{{\cal H}}
\def\cS{{\cal S}}

\def\cI{{\cal I}}

\def\cR{{\cal R}}

\def\cE{{\cal E}}

\DeclareMathOperator{\Tr}{Tr}
\DeclareMathOperator{\swap}{SWAP}
\usepackage{dsfont}
\def\I{\mathds{1}}

\begin{document}

\title{Popescu-Rohrlich box implementation in general probabilistic theory of processes}
\author{Martin Pl\'{a}vala}
\ead{martin.plavala@mat.savba.sk}
\address{Mathematical Institute, Slovak Academy of Sciences,\v Stef\'anikova 49, 814 73, Bratislava, Slovakia}
\address{Naturwissenschaftlich-Technische  Fakult\"{a}t, Universit\"{a}t Siegen, 57068 Siegen, Germany}

\author{M\'{a}rio Ziman}
\ead{mario.ziman@savba.sk}
\address{Institute of Physics, Slovak Academy of Sciences, D\'ubravsk\'a cesta 9, 845 11 Bratislava, Slovakia}
\address{Faculty of Informatics, Masaryk University, Botanick\'a 68a, 602 00 Brno, Czech Republic}

\begin{abstract}
It is shown that Popescu-Rohrlich nonlocal boxes (beating the Tsirelson bound for Bell inequality) do exist in the existing structures of both quantum and classical theory. In particular, we design an explicit example of measure-and-prepare nonlocal (but no-signaling) channel being the realization of nonlocal and no-signaling Popescu-Rohrlich box within the generalized probabilistic theory of processes. Further we present a post-selection-based spatially non-local implementation and show it does not require truly quantum resources, hence, improving the previously known results. Interpretation and potential (spatially non-local) simulation of this form of process nonlocality and the protocol is discussed.
\end{abstract}

\begin{keyword}
Bell inequalities \sep CHSH inequality \sep PR boxes \sep quantum channels
\end{keyword}

\maketitle

\section{Introduction}
The spooky paradox of Einstein, Podolski and Rosen, challenging the completeness of the mathematical framework of quantum theory, represents one of the deepest foundational puzzle of quantum theory. Quantum spookiness was discovered in 1935 \cite{EinsteinPodolskyRosen-paradox} and it took thirty years until this paradox has been resolved by John Bell in Ref.~\cite{Bell-ineq}. He has realized that theories compatible with the concept of local realism (used to create the paradox) necessarily satisfy certain inequalities, nowadays known as Bell inequalities. He has shown that quantum theory violates them and the phenomenon of quantum (Bell) nonlocality \cite{BrunnerCavalcantiPironioScaraniWehner-review} has been discovered. Since then the mathematical origin of the EPR paradox was clarified and nonlocality-enabled quantum technologies have been developed, however, its intuitive spookiness remains (see for instance Refs.~\cite{Peres-QT, WernerWolf-BellIneq}).

It is know that incompatibility is intimately related with the phenomenon of nonlocality \cite{WolfPerezgarciaFernandez-measIncomp, BuschHeinosaariSchultzStevens-compatibility, StevensBusch-CHSHIncomp, HeinosaariMiyaderaZiman-compatibility, FilippovHeinosaariLeppajarvi-compat, JencovaPlavala-maxInc}. Indeed, both the entangled states and the incompatible measurements are necessary for the demonstration of Bell nonlocality. The question of incompatibility of observables of processes was addressed recently in Ref. \cite{SedlakReitznerChiribellaZiman-compatibility}, where it was show that incompatibility of process measurements is both qualitatively and quantitatively different from incompatibility of state observables. In particular, unlike for quantum state measurements, for quantum process observables the maximum incompatibility degree is attained already for qubit systems. It was exactly this observation that has driven our curiosity to investigate the CHSH settings within the framework of general probabilistic theory of quantum processes (GPTQP).

Since quantum processes are represented by Choi matrices forming a subset density operators, one may think that whatever phenomena we observe for channels could be simulated by quantum states. However, as the study on process observables incompatibility shows, the reality is different. In this work we will demonstrate that also the phenomenon of non-locality is different. The paper is organized as follows: In the Section II we will define different classes of nonlocal boxes. The general probabilistic theory of quantum processes is described in Section III. Section IV is devoted to maximization of the violation of the Bell-CHSH inequality and we obtain results similar to \cite{BeckmanGottesmanNielsenPreskill-channels, HobanSainz-channels, Crepeau-CHSH}. In Section V we describe probabilistic spatial implementation of the CHSH test using only classical shared randomness and post-selection. Section VI is left for conceptual discussion of the achieved results.

\section{Nonlocal boxes}
In general, the paradigm of nonlocality requires the existence of spatially separated experimental facilities. It is common to name these local experimental stations as Alice and Bob. It is assumed that both can choose independently (of each other) their local experimental setups. In case of CHSH inequalities \cite{ClauserHorneShimonyHolt-CHSH} each of them is selecting between a pair of two-outcome measurement setups ($X\in\{A,A^\prime\}$ for Alice and $Y\in\{B,B^\prime\}$ for Bob). In particular, the outcomes are labeled as $a,a^\prime,b,b^\prime$, respectively, and takes values $\pm 1$. The whole setup is described by a family of conditional probabilities $P(x,y|X,Y)$, i.e. by 16 numbers satisfying the elementary probability constraints $0\leq P(x,y|X,Y)\leq 1$ and $\sum_{x,y}P(x,y|X,Y)=1$ for all $x,y=\pm 1$, $X\in\{A,A^\prime\}$ and $Y\in\{B,B^\prime\}$. We say these conditional probabilities describe a \emph{bipartite nonlocal box}, because the experiment can be rephrased as a black-box device with separated inputs $X,Y$ and outputs $x,y$ on Alice's and Bob's sides, respectively.

In a special case, when such nonlocal box can be simulated by means of local conditional probability distributions and exchange of classical information, we say the nonlocal box is of local hidden variable type, i.e. \emph{LHV box}. By definition it means that
\begin{equation*}
P(x,y|X,Y)=\sum_\lambda \pi(\lambda) P(x|X,\lambda) P(y|Y,\lambda)
\end{equation*}
for (local) conditional probabilities $P(x|X,\lambda)$, $P(y|Y,\lambda)$ and some probability distribution $\pi(\lambda)$ (called \emph{local hidden variable model} over the (local hidden) parameter $\lambda$. Clearly, the local hidden variable model $\pi(\lambda)$ represents the resource of (classical) randomness pre-shared between Alice and Bob prior to experiment. Such shared randomness is the physical implementation of LHV boxes. It is important to stress that choices of $X$ and $Y$ are independent of the local hidden variable $\lambda$.

By definition, LHV boxes satisfy all Bell inequalities \cite{Fine-BellIneq, Peres-BellIneq, WernerWolf-AllMultiparBellIneq}, thus, do not exhibit the phenomenon of quantum nonlocality. In particular, let us fix $X$ and $Y$ and denote by  $\langle X\otimes Y\rangle=\sum_{x,y=\pm 1} xy P(x,y|X,Y)$ the expectation value of their joint measurement performed by Alice and Bob, respectively. Since for all choices of $a,a^\prime,b,b^\prime=\pm 1$ the identity $a(b+b^\prime)+a^\prime(b-b^\prime)=\pm 2$ holds, it follows that for any LHV box $|\langle A\otimes B\rangle + \langle A\otimes B^\prime\rangle + \langle A^\prime\otimes B\rangle - \langle A^\prime\otimes B^\prime\rangle | \leq 2$. This inequality is known as CHSH-Bell inequality \cite{ClauserHorneShimonyHolt-CHSH}. 

The phenomenon of quantum nonlocality is recovered when the shared randomness is replaced by the shared bipartite quantum state that is accessed independently by Alice and Bob who performs measurements on their parts. \emph{Quantum (state) nonlocal boxes} are inducing the conditional probabilities through Born formula $P(x,y|X,Y)=\Tr (\varrho (A_x\otimes B_y))$, where $\varrho$ is some density operator (representing a joint state of a pair of quantum systems) and $A_x,B_y$ are effects (positive operators smaller than identity operator)  describing the probabilities of the outcomes observed by Alice and Bob, respectively. It is well known that many of quantum nonlocal boxes are violating CHSH inequality and that for any choice of the state and the measurements $|\langle A\otimes B\rangle + \langle A\otimes B^\prime\rangle +\langle A^\prime\otimes B\rangle - \langle A^\prime\otimes B^\prime\rangle | \leq 2\sqrt{2}$. This quantum limitation is known as Tsirelson bound \cite{Cirelson-bound}.

In their seminal work \cite{PopescuRohrlich-PRbox, Popescu-nonlocal} Popescu and Rohrlich questioned the conceptual and operational origin of quantum correlations and, especially, of the existence of Tsirelson bound. It is straightforward to verify that the no-signaling maximum is equal to the algebraic maximum of CHSH formula, which is four. Popescu and Rohrlich were wondering why this value of nonlocality is not achieved by quantum states. Extending CHSH framework they have identified the concept of nonlocal boxes and motivated the study of so-called general probabilistic theories (GPT) \cite{Barrett-GPTinformation, ChiribellaDArianoPerinotti-GPTpurification}. In particular, they have introduced a family of no-signaling nonlocal boxes characterized by the conditions
\begin{equation*}
P(x|X)=\sum_y P(x,y|X,Y)=\sum_{y^\prime} P(x,y^\prime|X,Y^\prime)
\end{equation*}
for any $X,Y,Y^\prime$ and
\begin{equation*}
P(y|Y)=\sum_x P(x,y|X,Y)=\sum_{x^\prime} P(x^\prime,y|X,Y)
\end{equation*}
for any $X,X^\prime,Y$. In other words, the marginal distributions $P(x|X)$, $P(y|Y)$ are independent of the choices of $Y$ and $X$, respectively. Such no-signaling restriction ensures that measurements performed in the local laboratories are not influencing each other.

Popescu and Rohrlich gave an example of no-signaling nonlocal box (coined as PR box) violating Tsirelson bound for CHSH inequality and achieving the algebraic maximum. Relabeling the outcome values to $\{0,1\}$ and denoting non-dashed and dashed observables as choices $0$ and $1$, respectively, we set for all variables the same value space $x,y,X,Y\in\{0,1\}$. Using such notation the conditional probabilities for PR box can be compactly expressed as $P_{\rm PR}(x,y|X,Y)=1/2$ if $x\oplus y=XY$ and vanishes otherwise.

Recently, it was realized that PR box would represent a resource both qualitatively and quantitatively stronger than any quantum nonlocal box \cite{Barrett-GPTinformation, CerfGisinMassarPopescu-simEnt, BrunnerCavalcantiPironioScaraniWehner-review, QuekShor-superquantChan}. For example \cite{BuhrmanChristandlUngerWehnerWinter-crypt, BrassardBuhrmanLindenMethotTappUnger-communication, vanDam-communication}, PR box would make any communication trivial from the complexity point of view. However, such device is purely hypothetical and currently no known existing realizations. Several proposals and attempts have been made to realize PR box in laboratories \cite{Cabello-BellIneq, MarcovitchReznikBenniVaidman-PRbox, ChuZongYangPanCao-PRboxSimulation}. All these constructions are employing post-processing of quantum measurements to achieve the desired goal. In this work we will present qualitatively different PR box implementations. In particular, we will show that a general probabilistic theory in which channels are playing the role of states allows for maximal violation of CHSH inequality. Moreover we will present a protocol based on post-selection and classical shared randomness, that implements the PR box correlations without the need for quantum systems at all.

\section{General probabilistic theory of quantum processes}
When designing a general probabilistic theory (GPT) we can follow the following algorithm. First, we specify a convex set playing the role of states (density operators in quantum theory), i.e. mathematical representation identified with the set of preparations of the object of experiments. Probabilities of outcomes of measurements are then naturally identified with positive affine (thus convex structure preserving) functions, i.e. affine maps from the convex set into the interval $[0,1]$. Such functions are usually called effects.

Further, we need to give some mathematical meaning to systems composed of more than one object (being the tensor product for quantum theory). There is a freedom in the definition of a suitable tensor product and a particular choice determines the features of the theory. However, the discussion of these consequences is not relevant for the purposes of this work, the only important fact is that the chosen tensor product must contain all separable states and only no-signaling states. There are several ways to express nonlocality of bipartite systems. One may e.g. consider spatial nonlocality, that is a bipartite system is nonlocal if its parts are separated by some distance or temporal nonlocality, that is when objects are separated by some time period. In our calculations we will use Bell nonlocality. We will say that a bipartite state is Bell nonlocal if it violates some Bell inequality. We will be concerned only with CHSH Bell inequality.

We will consider as a state space the set of quantum channels (quantum processes), i.e. the set of completely positive trace preserving linear maps mapping density operators to density operators. Quantum channels on a quantum system identified with $d$-dimensional complex Hilbert space $\cH_d$ are represented by their Choi matrices \cite{Choi-isomorph}, that are density operators $\Phi$ on $\cH_d\otimes\cH_d$ satisfying the normalization constraint $d \Tr_2 (\Phi)= \I$, where $\Tr_2$ denotes the partial trace and $\I$ denotes the identity matrix. In particular, $\Phi=(\cE\otimes\cI)[\omega_+]$, where $\omega_+=d^{-1}\sum_{jk} \ket{j\otimes j}\bra{k\otimes k}$ is the maximally entangled state, $\cE$ is the channel that corresponds to the Choi matrix $\Phi$ and $\cI$ denotes the identity map.

Therefore, for GPT of quantum processes (GPTQP) the convex set of states is identified with the set of Choi matrices $\Phi$. The process effects are then represented by functionals $f(\Phi) = \Tr (\Phi F)$, where $F$ is a positive operator describing probability of an outcome of an experiment addressing the properties of quantum channels. In general, the assignment of probabilities to experimental outcomes consists of preparation of an input state $\varrho$, application of the channel $(\cE\otimes\cI)[\varrho]$ and of applying some effect $0 \leq E \leq \I$, i.e. of computing $\Tr( (\cE\otimes\cI)[\varrho] E )$. The process effect $F$ captures the relevant characteristics of $\varrho$ and $E$. In particular, $F=(\cI\otimes\cR^*_\varrho)[E]$, where $\cR_\varrho$ is defined via the identity $(\cI\otimes \cR_\varrho)[\omega_+]=\varrho$ and $\cR^*_\varrho$ denotes the associated dual map \cite{Ziman-ppovm, ChiribellaDArianoPerinotti-PPOVM}, so we have $ \Tr (\Phi F) = \Tr( (\cE\otimes\cI)[\varrho] E )$. It is important to realize that the whole procedure of inputting a state and measuring the outcome is a local process effect $F$ in the framework of GPTQP, the application of the effect $E$ to $(\cE\otimes\cI)[\varrho]$ is just a part of the assignment of probability to the channel $\Phi$ (which is a state in GPTQP).

In order to discuss the non-locality phenomena for  GPTQP we need to introduce the concept of bipartite systems and local measurements. Naturally, these bipartite states coincide with bipartite channels defined on the tensor product of the underlying Hilbert spaces, i.e. $\cE:\cS(\cH_{AB})\to\cS(\cH_{AB})$, where $\cH_{AB}=\cH_A\otimes\cH_B$ is the joint Hilbert space composed of subsystems $A$ and $B$. In accordance with the single-partite case the associated Choi-Jamiolkowski operators $\Phi=\cE\otimes\cI_{AB}(\omega_+)$ form a subset of density operators $\cS(\cH_{AB}\otimes\cH_{AB})$ satisfying the constraint $d_{AB}{\rm tr}_2{\Phi}=I_{AB}$, where $d_{AB}$ is the dimension of $\cH_A\otimes\cH_B$ and $I_{AB}$ denotes the identity operator on this Hilbert space.  Local measurements on subsystem $A$ are described by effects of the form $F\otimes I_{B}\otimes I_B$, where $F$ is positive operator on $\cH_A\otimes\cH_A$ satisfying $F\leq I_A\otimes\varrho$ with $\varrho\in\cS(\cH_A)$. Analogously one identifies local measurements on subsystem $B$. Measurements performed on subsystem $A$ and subsystems $B$ individually and recording outcomes $a$ and $b$, respectively, consists of effects of the form $X_a\otimes Y_b$. Unlike for states, the no-signaling condition $p(a|X,Y)=p(a|X)$ does not hold in general, because $\sum_b {\rm tr}[\Phi(X_a\otimes Y_b)]={\rm tr}[\Phi (X_a\otimes I\otimes\varrho)]$ clearly depends on the normalization of the local process observable $Y$ on the subsystem $B$. It follows that nonlocal channels might be (in a sense naturally) signaling.

For example, consider a nonlocal unitary channel swapping states of the subsystems $A$ and $B$. It is straightforward to see that changing input states on $B$ (being part of the choice of local observable) changes the statistics of the measurement on the subsystem $A$, because the local measurements on $A$ are actually implemented on the input of $B$. In particular, setting $\Omega_+=\omega_+^{AA^\prime}\otimes\omega_+^{BB^\prime}$ the Choi operator of the swap unitary channel is $\Phi_{\rm SWAP}=\omega_+^{BA^\prime}\otimes\omega_+^{AB^\prime}$. A straightforward calculation gives
\begin{align*}
p(a|X,Y) &= \sum_b \Tr (\Phi_{\swap} X_a^{AA^\prime}\otimes Y_b^{BB^\prime}) \\
&= \frac{1}{d^2} \sum_{jk} \varrho_{jk} \Tr (X_a(\ket{j}\bra{k}\otimes \I ) ) \,,
\end{align*}
where $\varrho_{jk}= \Tr ( \ket{j} \bra{k} \varrho )$ and $\sum_{b}Y_b = \I_B \otimes \varrho$. Let us stress that $p(a|X,Y)$ depends on the normalization of the measurement $Y$, therefore we can conclude that the swap channel is signaling.

In summary, the GPTQP identifies the state space with the set of all Choi matrices of quantum channels. The role of process effects is played by positive operators $0 \leq F\leq \I \otimes \varrho$, for some density operator $\varrho$. The process observable is a collection of process effects $F_j$ satisfying the normalization $\sum_j F_j = I\otimes\varrho$. The composition of systems is described by the standard tensor product of the accompanying operator Hilbert spaces. Let us note that the abstract process-matrix formalism introduced in Ref.~\cite{OreshkovCostaBrukner-processMatrix} is generalizing the operationally motivated formalism of GPTQP.

\section{Channel-based CHSH no-signaling nonlocality}
Although signaling nonlocal channels exist in the framework of GPTQP \cite{BisioPerinotti-higherOrder}, from now on we will be interested only in the non-signaling nonlocal boxes and in the violations of the CHSH inequality that we can realize using the no-signaling nonlocal boxes.

Nonlocal boxes induced by quantum processes within GPTQP are described by a resource process state $\Phi$ being associated with a bipartite channel $\cE$ and a specification of a pairs of two-valued process observables $A,A^\prime$ and $B,B^\prime$ used by Alice and Bob, respectively. Within the framework of CHSH we assume the outcomes of all the observables are taking values $\pm 1$. Consequently, such process-based nonlocal boxes lead to probabilities $P(\pm,\pm|A,B)=\Tr(\Phi A_\pm\otimes B_\pm)$. The expectation value for $X \otimes Y$ equals 

\begin{equation*}
\langle X\otimes Y\rangle_\Phi = \Tr ( \Phi (X_a\otimes Y_b) )= \sum_{a,b=\pm 1} ab P(a,b|X,Y)\,.
\end{equation*}
Our goal is to find the maximal value of the CHSH expression $\langle (A+A^\prime)\otimes B + (A-A^\prime)\otimes B^\prime\rangle_\Phi$ for choices of $\Phi$ and $A,A^\prime,B,B^\prime$. In what follows we will give an example achieving the value of 4 that is known to be the algebraic maximum.

Denote by $\ket{0},\ket{1}$ the vectors forming the computational basis of qubit Hilbert space $\cH_2$. Consider a two-qubit measure-and-prepare channel in which, each of the qubits is (independently) measured in the computational basis and subsequently (based on the observed outcomes $j,k\in\{0,1\}$) the two-qubits are prepared in a fixed two-qubit state. In particular,  $\xi_{\rm cor}=\frac{1}{2}(\ket{00}\bra{00}+\ket{11}\bra{11})$ if $jk=0$ and $\xi_{\rm acor}=\frac{1}{2}(\ket{01}\bra{01}+\ket{10}\bra{10})$ if $jk=1$. These output states describe the (classically) correlated and anticorrelated state of two qubits, respectively. In summary, the total action of such channel leads to the following state transformation

\begin{equation*}
\cE[\varrho]=(1-\kappa)\xi_{\rm cor}+\kappa\xi_{\rm acor}
\end{equation*}
where $\kappa=\varrho_{11,11}=\bra{11}\varrho\ket{11}$.

In more technical terms, this channel acts as follows $\cE[X]=({\rm tr}X-\kappa)\xi_{\rm corr}+\kappa\xi_{\rm acor}$, thus, Choi operator reads $\Phi=\xi^{AB}_{\kappa}\otimes I_{A^\prime B^\prime}$ where $\xi_\kappa=[(1-\kappa)\xi_{\rm corr}+\kappa\xi_{\rm acor}]$. Then

\begin{align*}
 p(a|X,Y) &= \Tr (\Phi(X_a^{AA^\prime}\otimes \I^B\otimes \varrho^{B^\prime})) \\
&= \Tr ((\xi_\kappa^{AB}\otimes \I^{A^\prime})(X_a^{AA^\prime}\otimes \I^B) \otimes \varrho^{B^\prime}) \\
&= \Tr ((\xi_\kappa^{AB}\otimes \I^{A^\prime})(X_a^{AA^\prime}\otimes \I^B)) \\
&= p(a|X)\,,
\end{align*}
thus, it is independent on the choice of measurement $Y$ on the subsystem B. In a similar fashion one can show that the same holds for the choice of the measurement $X$ on the subsystem A. Consequently, the considered channel $\cE$ is no-signaling.
   
In order to perform Bell like experiment we need to specify local process measurements of such nonlocal channel. Let us denote by $Z_\psi$ a process measurement, in which the quantum channel acts on the single qubit test state $\ket{\psi}$ and the output system is measured in $\sigma_z$ basis. Bare in mind that both inputting the testing state $\ket{\psi}$ and measuring in $\sigma_z$ basis are parts of the local process measurement. Moreover in the case that we will present, inputting the state $\ket{\psi}$ can be identified with pressing certain button on the PR box. This also shows that the process measurement is a local action. 

Suppose that on both sides one may perform either $Z_0$, or $Z_1$ process measurement, thus, the test states are $\ket{0}$, or $\ket{1}$, respectively. In the settings of Bell inequalities $X,Y\in\{Z_0,Z_1\}$ and $x,y=\{\pm 1\}$. We have everything we need to evaluate the conditional probabilities $P(x,y|X,Y)$ and, subsequently, evaluate the CHSH expression. Let us denote by $\cE(jk)=\cE(\ket{j}\bra{j}\otimes\ket{k}\bra{k})$. For measurement choices $Z_j,Z_k$ with $jk=0$ we obtain
\begin{align*}
P(+,+|Z_j,Z_k)&= \bra{00}\cE(jk)\ket{00} = \bra{00}\xi_{\rm cor}\ket{00} = \frac{1}{2} \\
P(+,-|Z_j,Z_k)&= \bra{01}\cE(jk)\ket{01} = \bra{01}\xi_{\rm cor}\ket{01} = 0 \\
P(-,+|Z_j,Z_k)&= \bra{10}\cE(jk)\ket{10} = \bra{10}\xi_{\rm cor}\ket{10} = 0 \\
P(-,-|Z_j,Z_k)&= \bra{11}\cE(jk)\ket{11} = \bra{11}\xi_{\rm cor}\ket{11} = \frac{1}{2}.
\end{align*}
Consequently,
\begin{equation*}
\langle Z_0\otimes Z_0 \rangle_\Phi= \langle Z_0\otimes Z_1 \rangle_\Phi= \langle Z_1\otimes Z_0 \rangle_\Phi=1\,,
\end{equation*}
For the remaining choice $Z_1, Z_1$ one finds
\begin{align*}
P(+,+|Z_1,Z_1)&= \bra{00}\cE(11)\ket{00} = \bra{00}\xi_{\rm acor}\ket{00} = 0 \\
P(+,-|Z_1,Z_1)&= \bra{01}\cE(11)\ket{01} = \bra{01}\xi_{\rm acor}\ket{01} = \frac{1}{2} \\
P(-,+|Z_1,Z_1)&= \bra{10}\cE(11)\ket{10} = \bra{10}\xi_{\rm acor}\ket{10} = \frac{1}{2} \\
P(-,-|Z_1,Z_1)&= \bra{11}\cE(11)\ket{11} = \bra{11}\xi_{\rm acor}\ket{11} = 0
\end{align*}
and $\langle Z_1\otimes Z_1 \rangle_\Phi=-1$. Therefore,
\begin{equation*}
\langle (Z_0+Z_1)\otimes Z_0 + (Z_0-Z_1)\otimes Z_1\rangle_\Phi = 4 \,.
\end{equation*}
In conclusion, this proves that the designed experiment achieves algebraic maximum of CHSH quantity, thus, overcoming the Tsirelson bound valid for quantum states and demonstrating the existence of PR box. Let us recall that the induced probabilities $P(x,y|X,Y)$ and the channel $\cE$ are indeed no-signaling.

\section{Protocol for probabilistic spatially non-local implementation
  of the nonlocal channel}
Although the employed channel $\cE$ is no-signaling, it is nonlocal.
Since we do not have access to long-range interaction, its spatial distribution is extremely limited. To implement the nonlocal channel $\Phi$ we are going to use a modified version of the protocol for probabilistic storing and retrieving of unitary channels \cite{SedlakBisioZiman-chanStoring}. The idea is as follows: in quantum teleportation there is a one in four chance that one does not have to do any corrections to the resulting state. In this case, we can first input one part of an entangled state into the channel (store the action) and later with some probability teleport the input state into the channel (retrieve the action).

\begin{figure}
\centering
\includegraphics[width=\linewidth]{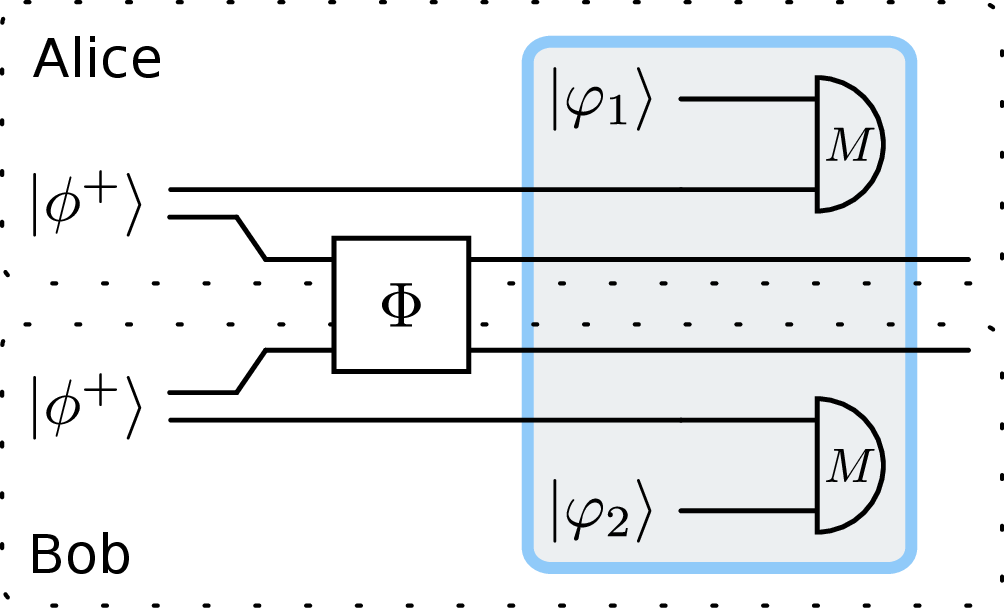}
\caption{The protocol for implementing a nonlocal channel $\Phi$ with inputs $\ket{\varphi_A}$ and $\ket{\varphi_B}$ using probabilistic channel storing. $\ket{\Phi^+}$ denote the Bell states, $M$ denotes the measurement in Bell basis.} \label{fig:circuit}
\end{figure}

To be exact, the protocol (depicted in Fig. \ref{fig:circuit}) consist of two parties, say Alice and Bob, both having in their possession a maximally entangled state $\ket{\phi^+} = \frac{1}{\sqrt{2}} \sum_{ij} \ket{i \otimes i}\bra{j \otimes j}$. In the first step (that corresponds to the preparation phase of the nonlocal resource), both Alice and Bob take one of the two particles forming their respective maximally entangled pair and input this particle into the channel $\Phi$ and they receive back the respective output. After the first step, the laboratories of Alice and Bob may be spatially separated, each holding a pair of particles representing (storing) the nonlocal resource and a Bell measurement device $M_{\rm Bell}$ performing measurement in the Bell basis $\ket{\phi^+}, \ket{\phi^-}, \ket{\psi^+}, \ket{\psi^-}$.

Let us note that the shared resource state 
$\varrho_{\cE} = (\cI \otimes \cE \otimes\cI)( \ket{\phi^+}\bra{\phi^+} \otimes \ket{\phi^+}\bra{\phi^+} )$ (with $\cI$ denotes the identity map) storing the action of $\cE$ reads

\begin{align*}
\varrho_{\cE} = \dfrac{1}{4} \big( &
\ket{0} \bra{0} \otimes \xi_{\rm cor} \otimes \ket{0} \bra{0} + \ket{0} \bra{0} \otimes \xi_{\rm cor} \otimes \ket{1} \bra{1} + \\
&\ket{1} \bra{1} \otimes \xi_{\rm cor} \otimes \ket{0} \bra{0} + \ket{1} \bra{1} \otimes \xi_{\rm acor} \otimes \ket{1} \bra{1}
\big).
\end{align*}
Let us stress that the state $\varrho_{\cE}$ is a classically correlated state and that it has zero quantum discord \cite{OllivierZurek-discord, AdessoCianciarusoBromley-discord, Vedral-discord}, hence, the storage creates purely classical correlations between the involved qubits. The protocol (in the forthcoming step) will work the same way if the classically correlated and uncorrelated states are replaced by entangled correlated and anticorrelated states, e.g. $\xi_{\rm cor} = \ket{\Phi^\pm}\bra{\Phi^\pm}$ and $\xi_{\rm acor} = \ket{\psi^\pm}\bra{\psi^\pm}$, respectively. But there is no need to involve entanglement.

One may object that presharing the state $\varrho_{\cE}$ is not a local procedure and requires communication. However, any entanglement-based protocol (teleportation, key distribution, etc.) that requires entangled states requires the parties to preshare the entangled state before running the protocol. 

In the second step Alice and Bob access the nonlocal box, i.e. both of them prepare input states $\ket{\varphi_A}$ and $\ket{\varphi_B}$ (both being either $\ket{0}$, or $\ket{1}$). Further they perform measurement $M_{\rm Bell}$ on their input state and on the remaining (unused) particle of the maximally entangled pair. If both of them observe the outcome $\ket{\phi^+}$, they are left with the output of the channel $\cE$ applied on $\ket{\varphi_A}$ and $\ket{\varphi_B}$. This output is stored in the remaining unmeasured qubits and they may continue with the CHSH test.

Let us stress that measuring the outcomes $\ket{\phi^-}$ corresponds to situation when Alice and Bob inputs are $\sigma_z$ rotated, however, since their inputs are either $\ket{0}$, or $\ket{1}$, they remain unchanged ($\sigma_z\ket{\varphi_A}=\ket{\varphi_A}$ and $\sigma_z\ket{\varphi_B}=\ket{\varphi_B}$). In other words also the simultaneous observations of outcome $\ket{\phi^-}$ or observation of any combinations of the outcomes $\ket{\phi^+}$ and $\ket{\phi^-}$ correspond to a successful run of the protocol.

In order to verify the implementation is successful Alice and Bob need to communicate whether they are successful on their sides, or not. In total, the protocol has $50\%$ success chance for each of the parties, hence, overall the protocol has $25\%$ chance of being successful. It is important to note that the transmitted classical bits are random and it is impossible to deduce from them the particular inputs of Alice and Bob. Once the success is confirmed they both can continue to perform the CHSH test, i.e. measure locally the remaining states by a randomly selected observable ($X$ and $Y$) and record the outcomes $\pm 1$.

\section{Discussion}
It is well known that PR boxes do exist as mathematical objects in artificial general probabilistic theories. In this work we have shown that PR boxes are accommodated within an existing formalism of quantum theory - the GPTQP, thus, PR boxes are not anymore purely conceptual objects appearing in thought experiments. Indeed, the above equations for conditional probabilities show explicitly that the channel $\Phi$ is a mathematically faithful realization of PR box.

A question that emerges is related to the interpretation of nonlocality. Unlike for the case of quantum states, it is not possible to separate parts of the channel and distribute them to distant places. In current physics, we are lacking long-distance interactions and therefore it is essentially forbidden to implement the PR box channel $\Phi$ in a spatially separated manner. However, the mathematical formalism of nonlocal boxes does not rely on the concept of space. The concept of locality can be understood purely in terms of subsystems, i.e. in our ability to address subsystems individually. Adopting such perspective we do not affect any mathematical statement regarding the nonlocality, thus, the existence of such nonlocality with respect to subsystems is of foundational relevance and interest. On the other side in such case the interpretation of nonlocal boxes as nonlocal objects in space is not necessarily applicable and this fact have potential consequences for communication-based applications, where the spatial separation between the participants is essential.

Consequently, for sake of clarification it turns out to be important to distinguish conceptually between Bell nonlocality and spatial nonlocality. Having such clarification in mind we may say that the introduced channel $\Phi$ is a faithful implementation of PR box not reflecting the spatial nonlocality. Following this interpretation, one may wonder to which extent the spatial nonlocality of $\Phi$ can be simulated. Similar question we are facing in entanglement swapping protocols \cite{ZukowskiZeilingerHorneEkert-entanglementSwap,PanBouwmeesterWeinfurterZeilinger-experimentEntalgementSwap}, where we are aiming to implement the swap gate between spatially separated systems. Again, such transformation is forbidden in universes without nonlocal interactions. However, having access to entanglement and communication such gate can be simulated (essentially by implementing the quantum teleportation protocol \cite{BennettBrassardCrepeauJozsaWootters-teleportation}). The drawback is that the gate implementation is not instantaneous. In fact, its realization is limited by the speed of communication, thus, by the mutual space distance between the systems. We have seen the same problem emerge in the proposed probabilistic protocol for implementing the nonlocal channel $\Phi$. Although the parties communicate only random bits to verify that the protocol has worked, the communication is still crucial. Yet it is an open question whether there do exist instantaneous implementations of the proposed nonlocal channel $\Phi$. The reason why they might exist is that the channel $\Phi$ is no-signaling channel, hence it does not transmit any information and it is not restricted by the speed of light. On the other side, this question is conceptually related to the principal universality of the Tsirelson's bound that would imply limitations on any implementation of nonlocal chanels. 

In practice, the proposed CHSH experiment with PR box can be seen as follows. Both Alice and Bob choose one random bit $j$ and $k$ (determining the input test state). If $jk = 0$, then (after the action of the channel $\Phi$) they observe perfectly correlated outcomes and if $jk = 1$ they found their results perfectly anticorrelated. Seeing the whole experiment from this perspective it is natural to ask what is quantum in this experiment? The answer may sound rather surprising. The channel $\Phi$ can be seen as purely classical channel acting on a pair of classical bits and all the results derived for $\Phi$ remain valid. This is clearly very unexpected result that PR-type nonlocality emerges already in the mathematical formalism of classical structures, namely for general probabilistic theory of classical processes. It is even more astonishing that the proposed probabilistic protocol for implementing the nonlocal channel $\Phi$ can be considered to be a classical protocol, although the motivation for it comes from quantum protocols.

In conclusion, we found that PR box can be implemented within general probabilistic theory of quantum and classical processes.  This implies that superquantum no-signaling correlations are not forbidden in our experiments, however, also that the spatial nonlocality should be distinguished from Bell nonlocality. Our observation induces a plethora of open questions and exciting foundational research directions. For example, the presented construction is based on measure-and-prepare channel that is in a sense classical.

One may wonder whether there are also some intrinsically non-classical channels maximizing the CHSH expression. More general characterization of superquantum nonlocal boxes will be addressed in future work.

\section*{Acknowledgments}
We thank the anonymous referee for very valuable feedback. This work was supported in part by research project APVV-18-0518 (OPTIQUTE) and COST Action CA15220. MP is thankful for the support by Grant VEGA 2/0142/20, by the grant of the Slovak Research and Development Agency under Contract No. APVV-16-0073, by the DFG and by the ERC (Consolidator Grant 683107/TempoQ). MZ acknowledges support of projects VEGA 2/0173/17 (MAXAP), GA\v{C}R No. GA16-22211S and MUNI/G/1211/2017 (GRUPIK).


\end{document}